\documentclass[fleqn,10pt]{SelfArx}

\usepackage{colortbl}
\usepackage{lipsum}
\usepackage{amsmath, amssymb}
\usepackage{graphicx}
\usepackage{booktabs}
\usepackage{caption}
\usepackage{subcaption}
\usepackage{float}
\usepackage{enumitem}
\usepackage{algorithm}
\usepackage{algpseudocode}
\usepackage{hyperref}
\usepackage{array}

\hypersetup{hidelinks,colorlinks=true,linkcolor=blue,citecolor=blue,urlcolor=blue}

\definecolor{color1}{RGB}{0,0,90}
\definecolor{color2}{RGB}{0,20,20}

\JournalInfo{Draft Version}
\Archive{}
\PaperTitle{Behind the Byline: A Large-Scale Study of Scientific Author Contributions
}

\Authors{Itai Assraf and Michael Fire}
\affiliation{\textsuperscript{1}\textit{Department of Data Engineering, Ben-Gurion University of the Negev, Israel}}
\affiliation{\textbf{Corresponding authors}: Itaias@post.bgu.ac.il and mickyfi@bgu.ac.il }

\Keywords{Scientometrics --- Science of Science  --- Authorship Attribution --- Contribution Mining --- Scholarly Metadata --- CRediT Taxonomy}
\newcommand{\keywordname}{Keywords}

\Abstract{Understanding how co-authors distribute credit is critical for accurately assessing scholarly collaboration. In this study, we uncover the implicit structures within scientific teamwork by systematically analyzing author contributions across a large corpus of research publications. We introduce a computational framework designed to convert free-text contribution statements into 14 standardized CRediT categories, identifying clear and consistent positional patterns in task assignments. By analyzing over 337,000 scientific articles from prominent sources such as PLOS One and Nature, we extracted and standardized more than 5.6 million author-task assignments corresponding to 1.58 million author mentions. Our analysis reveals substantial disparities in workload distribution. Notably, in small teams with three co-authors, the most engaged contributor performs over three times more tasks than the least engaged, a disparity that grows linearly with team size. This demonstrates a consistent pattern of central and peripheral roles within modern collaborative teams. Moreover, our analysis shows distinct positional biases in task allocation: technical responsibilities, such as software development and formal analysis, broadly fall to authors positioned earlier in the author list, whereas managerial tasks, including supervision and funding acquisition, increasingly concentrate among authors positioned toward the end. This gradient underscores a significant division of labor, where early-listed authors mainly undertake most hands-on activities. In contrast, senior authors mostly assume roles involving leadership and oversight. Our findings highlight the structured and hierarchical organization within scholarly collaborations, providing deeper insights into the specific roles and dynamics that govern academic teamwork.
}

\begin{document}

\flushbottom
\maketitle
\thispagestyle{empty}


\section{Introduction}
\addcontentsline{toc}{section}{Introduction}
In academic publishing, allocating credit among authors is a critical practice that acknowledges each contributor's role in the research and writing process \cite{icmje2024guidelines}.
Typically, the authorship order reflects the level of contribution: the first author is usually the primary contributor. In contrast, the last author often serves as the senior researcher or principal investigator~\cite{wren2007authorship}.
This division of credit is essential for recognizing individual input, ensuring accountability across different aspects of the work, and encouraging collaborative efforts among researchers.
Transparent credit allocation plays a vital role in academic careers by influencing professional recognition, career advancement, and future research opportunities~\cite{smith1997authorship}.
Thus, clear guidelines and ethical authorship practices are necessary to prevent disputes and promote fairness in the scholarly community \cite{marusic2011responsible}.

In many leading journals and conferences, one of the publication requirements includes composing a paragraph that outlines each author’s specific contribution to the article and the overall project.
This practice raises several questions regarding the meaning of authorship order and the diversity of roles, especially in large-scale collaborations \cite{allen2014credit}.

\begin{figure}[htbp]
    \centering
    \includegraphics[width=1\linewidth]{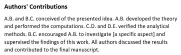}
    \caption{An example of the Authors' Contribution section.}
    \label{fig:example}
\end{figure}

Research into author contribution analysis has progressed in recent years, with multiple tools developed to extract structured information from scientific literature.
While systems like CERMINE \cite{tkaczyk2015cermine}, GROBID \cite{lopez2009grobid}, PDFX \cite{constantin2013pdfx}, and ParsCit \cite{councill2008parscit} focus on extracting bibliographic metadata, few are specifically designed to interpret or analyze individual author contributions.
Studies such as~\cite{brand2015beyond} have explored the correlation between authorship order and contribution type. Nevertheless, challenges like inconsistent authorship conventions and varied name representations remain unresolved.
Initiatives like the CRediT taxonomy, introduced by platforms such as PLOS One  and Nature, attempt to standardize author roles across publications \cite{allen2014credit}, but comprehensive, scalable solutions remain limited \cite{tkaczyk2019naiverole}.

\begin{figure}[htbp
]
    \centering
    \includegraphics[width=0.8\linewidth]{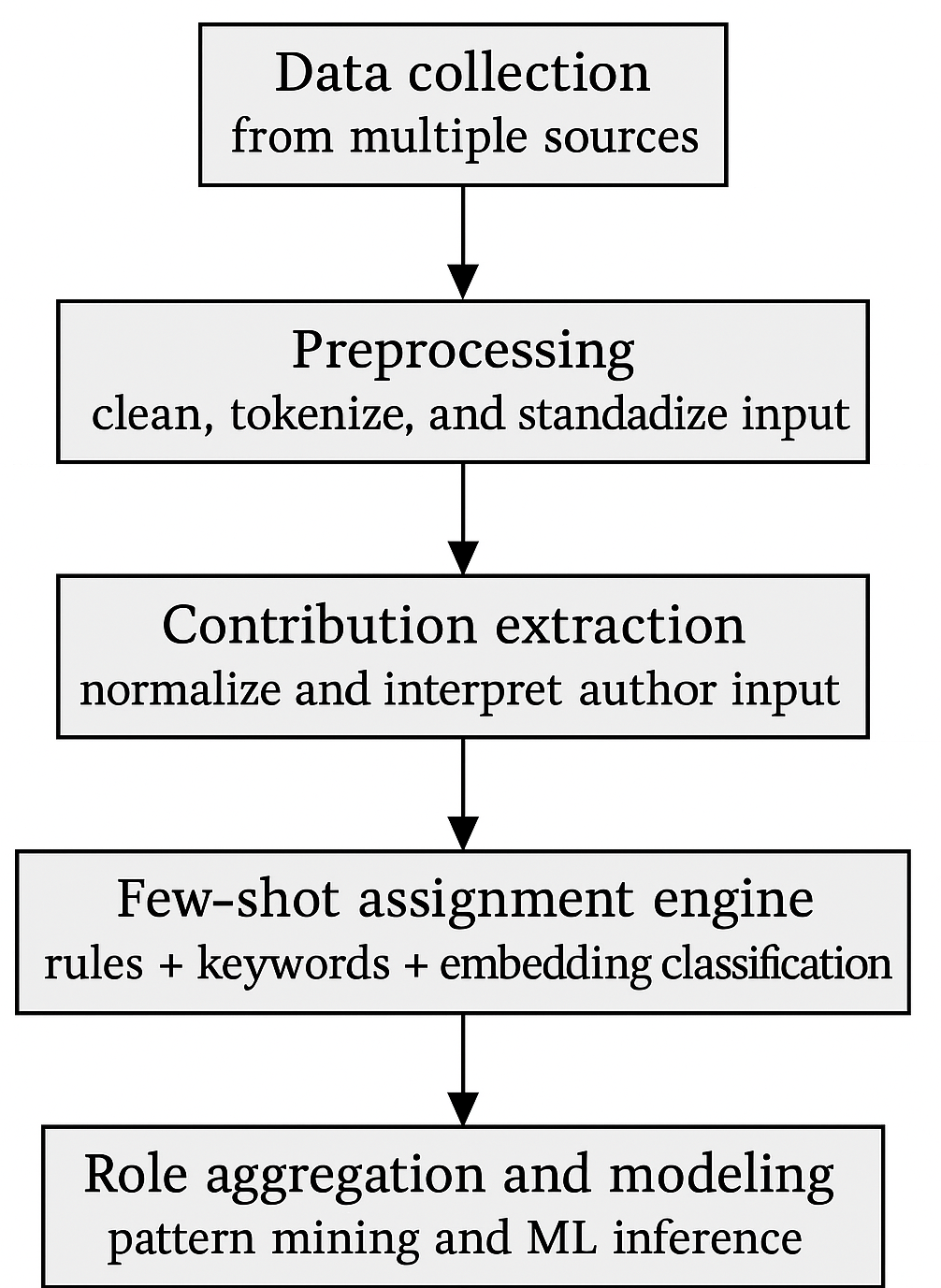}
    \vspace{1.5em}
    \caption{Overview of the methodology pipeline used to extract and analyze author contributions.}
    \label{fig:pipeline}
\end{figure}

To address these challenges, we developed a comprehensive framework for extracting, standardizing, and classifying author contributions across large-scale scientific corpora. As illustrated in Figure~\ref{fig:pipeline}, our framework includes structured data collection from multiple repositories, followed by text preprocessing, author name disambiguation, and classification of contribution statements into 14 predefined CRediT task categories. We incorporated the Apriori algorithm~\cite{agrawal1994fast} to complement this classification to uncover frequently co-occurring contribution patterns. This integration of pattern mining enables a structured understanding of how tasks are commonly bundled within collaborative teams. Our unified framework enables consistent extraction and classification of author contributions from diverse publication sources, laying the groundwork for large-scale quantitative analysis of contribution patterns and authorship dynamics.

To implement the classification step within this framework, we applied the following multi-stage assignment procedure.
We analyzed the extracted contribution text and assigned contributions to predefined task categories, such as \textit{writing – review editing}, \textit{methodology}, or \textit{data curation}. 
This classification was conducted using a hybrid assignment engine that integrates rule-based logic, keyword matching, and embedding-based few-shot learning. Specifically, we first applied deterministic rules to handle ambiguous or frequently conflated categories—such as distinguishing between writing – review editing and writing – original draft based on the presence of the word “draft.” We then used curated keyword lists for each of the 14 CRediT categories to match common task expressions. If neither rule nor keyword heuristics provided a confident assignment, we computed a semantic similarity between the author contribution sentence and a set of prototypical examples per category, using a pre-trained SentenceTransformer model ~\cite{reimers2019sentence}. Each category was represented by a vector derived from a few labeled examples and its domain-specific keywords. This few-shot representation enabled accurate matching of semantically similar but lexically varied contributions, especially in free-form author statements.

For example, given an article stating, "A.B. performed the data analysis, C.D. conceptualized the study, and E.F. wrote the initial draft," the algorithm matches initials to full names from the metadata. It identifies keywords like "data analysis," "conceptualized," and "draft" to assign A.B. to \textit{formal analysis}, C.D. to \textit{conceptualization}, and E.F. to \textit{writing – original draft}.

To evaluate the robustness and scalability of our classification system (see Section~\ref{sec:contrib_classification}), we applied it to a large-scale dataset containing over 337,000 scientific articles from sources such as PLOS One  and Nature. This extensive application allowed us to extract and standardize over 5.6 million author-task assignments corresponding to 1.58 million author mentions. These contributions were mapped into 14 CRediT-defined categories, enabling consistent comparison across disciplines and collaboration sizes.

Moreover, we performed a statistical dataset analysis after applying H-Contrib (Algorithm~\ref{alg:hcontrib}), a hybrid classification pipeline developed in this study. H-Contrib integrates rule-based keyword matching with embedding-based semantic similarity to assign free-text author statements to one of 14 standardized CRediT categories.
 
The algorithm revealed that \textit{writing – review editing} was the most frequent contribution category, appearing in 16\% of contributions. In contrast, \textit{software development} appeared only in 2\%. Using the Apriori algorithm, we identified eight frequent co-occurrence patterns of task categories. The most prominent combinations included \textit{writing – review editing} alongside \textit{formal analysis} (30\%), \textit{conceptualization} (29\%), and \textit{investigation} (28\%), with all three occurring together in 18\% of cases.

Lastly, our analysis of author positioning revealed that authors listed earlier tend to contribute to more tasks than those listed later. For example, across publications with 2 to 20 co-authors, the first author contributed on average 1.5 times more than the second author and 1.65 times more than the last author. Additionally, we developed a classification model to predict task assignments based on metadata features such as author position and team size. The model achieved an average accuracy of 0.84 across the 14 task categories, outperforming a logistic regression baseline, which reached 0.81.

\begin{figure}[htbp]
    \centering
    \includegraphics[width=1\linewidth]{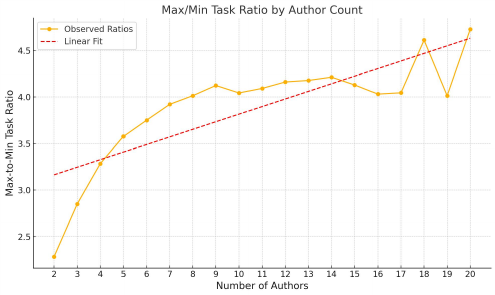}
    \caption{Relationship between the number of authors and the ratio of tasks performed by the most active contributor versus the least active. The linear trend suggests that larger teams exhibit greater contribution level disparity. A linear regression fitted to this trend yielded the equation: \ensuremath{\mathrm{Ratio}_{\mathrm{max/min}} = 3.000 + 0.082 \cdot \mathrm{TeamSize}}.}
    \label{fig:min max ratio tasks}
\end{figure}

In addition, we found that the disparity between contributors tends to increase with team size. Specifically, Figure~\ref{fig:min max ratio tasks} shows that when analyzing the ratio between the contributors most involved and the least involved in each article, the average ratio increased from 2.28 for teams with two authors to 
3.28 (5 authors), 4.13 (10 authors), and up to nearly 4.73 in teams with 20 authors, as illustrated in Figure~\ref{fig:min max ratio tasks}.

    Our work makes the following key contributions:

\begin{itemize}
    \item We propose a hybrid classification pipeline that combines rule-based heuristics with semantic similarity techniques to map free-text author statements into the standardized CRediT taxonomy.
    \item We apply this method to an extensive dataset of more than 337,000 scientific articles, enabling large-scale analysis of author behavior across disciplines and publication venues.
    \item We empirically demonstrate consistent patterns in the relationship between author position, team size, and contribution type,
    providing quantitative validation for long-held assumptions about academic collaboration.
    \item We develop and evaluate a predictive model capable of inferring author roles from simple metadata, achieving high accuracy across 14 contribution categories.
    \item We publicly release a large, structured dataset of author contributions to support further research on scientific collaboration and credit attribution.
\end{itemize}

The remainder of this paper is organized as follows: Section~\ref{sec:related_work} reviews existing work on author contribution modeling and scholarly metadata. Section~\ref{sec:methodology} describes our dataset, preprocessing steps, and classification approach. Section~\ref{sec:results} presents our empirical findings and model evaluation. Section~\ref{sec:discussion} presents key insights and outlines the main limitations of the study. Finally, Section~\ref{sec:conclusion} concludes the paper and outlines future research directions.

\section{Related Work}
\label{sec:related_work}
The field of information extraction from scientific literature has witnessed significant progress in recent years, driven by the need to structure, index, and analyze the rapidly growing volume of academic publications. Various systems and frameworks have been developed to automate the extraction of structured metadata and bibliographic elements from scholarly articles. Notable examples include \textit{CERMINE} \cite{tkaczyk2015cermine}, \textit{GROBID} \cite{lopez2009grobid}, \textit{PDFX} \cite{constantin2013pdfx}, and \textit{ParsCit} \cite{councill2008parscit}. These tools play a crucial role  in extracting elements such as titles, abstracts, references, author names, and affiliations, thereby enabling improved search, indexing, and citation analysis.

However, despite their utility, these systems are not designed to extract semantically rich information about the contributions of authors. This constitutes a significant gap in the current landscape of information extraction, as identifying and understanding each author's contribution is critical for evaluating scientific impact and fairly assigning credit.

One of the core difficulties in this area is the absence of universal standards for defining and reporting author contributions. While guidelines exist—such as those published by the \textit{International Committee of Medical Journal Editors (ICMJE)} \cite{icmje2024guidelines}—adherence remains voluntary, and interpretations vary across disciplines and journals. As a result, inconsistencies in authorship practices are widespread, undermining transparency and accountability in collaborative research \cite{icmje2024guidelines}.

Another technical challenge lies in matching author names and initials within contribution statements. Authors often use inconsistent formatting for initials, including or omitting middle initials, using periods or not, and varying the order of names. These discrepancies complicate linking initials in the contribution text with full author names in the metadata, making automated attribution prone to error.

To mitigate these challenges, some scholarly platforms— such as \textit{PLOS One} and \textit{Nature}—have implemented structured author contribution statements based on the standardized \textit{CRediT taxonomy} \cite{allen2014credit}. These systems prompt authors to specify their roles from a predefined set of categories, such as \textit{conceptualization}, \textit{methodology}, \textit{investigation}, \textit{writing – original draft}, \textit{writing- review editing}, \textit{project administration}, etc. This standardization improves transparency and facilitates downstream processing of author roles.
Several studies have explored the link between author order and contribution roles, providing valuable groundwork. Nonetheless, to the best of our knowledge, large-scale analyses have been limited, mainly due to the difficulty of aligning free-text statements with individual authors. Some efforts, such as \textit{NaïveRole} \cite{tkaczyk2019naiverole}, use rule-based approaches to associate textual cues with contribution categories. These approaches allow for semi-automated extraction of author roles, providing a foundation for contribution-aware author profiling.

Recent advancements in semantic similarity-based text classification have introduced methods that leverage sentence embeddings to map input statements and class labels into a shared vector space. Approaches such as SimCSE~\cite{gao2021simcse} and Lbl2Vec~\cite{schopf2021lbl2vec} have demonstrated improved classification performance, particularly in low-resource scenarios, by enabling inference based on proximity in embedding space. These methods align with our use of few-shot learning and embedding-based semantic comparison to classify author contributions, particularly in cases where keyword heuristics are insufficient.

\subsection{Author Contribution Analysis}
\label{sec:Author Contribution Analysis}
A growing body of literature has explicitly focused on understanding how scholarly tasks are distributed among authors and how contribution statements reflect these divisions. These works can be broadly divided into empirical studies, ethical analyses, and computational frameworks.

A notable large-scale empirical effort in this area was conducted by Corrêa Jr. et al.~\cite{correa2017patterns}, who analyzed thousands of contribution statements from PLOS ONE to investigate how work is distributed among authors. They proposed three dominant patterns of contribution allocation: a monotonic decline in contributions from first to last author; a lead-author-centric model in which the first author undertakes most of the work; and a U-shaped pattern where both the first and last authors contribute substantially, while middle authors play more specialized or technical roles. These findings provided an early taxonomy of contribution styles in multi-author teams. They highlighted how the author's position correlates with the contribution scope.

Building on this, Larivière, Pontille, and Sugimoto \cite{lariviere2021division} conducted a broader analysis using over 30,000 articles from PLOS journals that employed the structured CRediT taxonomy. Their study explored how labor is divided across contributors using a standardized 14-role framework, allowing for finer-grained comparisons across disciplines and author positions. Notably, they confirmed the prevalence of core-periphery team structures, where first and last authors often shared primary conceptual and writing duties, and middle authors contributed to more limited roles such as data collection or analysis. Additionally, they uncovered a gendered division of labor: women were more frequently assigned to supportive or technical roles, whereas men were overrepresented in leadership-oriented categories.

A related line of inquiry was pursued by Macaluso et al. \cite{macaluso2016gender}, who conducted a gender-focused analysis of contributorship across PLOS ONE articles. Their findings indicated consistent disparities in role attribution by gender. Women were significantly more likely to be credited with performing experiments and data curation tasks. At the same time, men more often appeared in supervisory or conceptual roles. These patterns reinforced the concern that authorship order and even contribution statements may mask underlying inequalities in team science.

Sauermann and Haeussler \cite{sauermann2017authorship} added further nuance by combining manual annotation of 12,000 contribution sections with survey responses from researchers. Their study revealed a systematic disconnect between author order and actual labor: middle authors sometimes performed central tasks, while senior figures were occasionally included with little documented input, highlighting the persistence of “honorary” or “guest” authorship. Their survey also revealed ambivalent community attitudes toward contribution disclosures, with some researchers expressing skepticism about the ethics of assigning equal credit.

Ueda et al. \cite{ueda2021authororder} focused more narrowly on the statistical relationship between author position and reported task. Using a dataset of 576 papers, they showed that while author order remains loosely indicative of workload, it cannot reliably serve as a proxy for contribution type. Their work emphasized the need for more robust, standardized systems of contribution tracking to replace inference based on sequence alone.

Beyond role allocation, several studies have turned attention to distortions in authorship practices arising from the misuse of bibliometric incentives. Fire and Guestrin \cite{fire2019goodhart} examined over 120 million publications and revealed how optimization of metrics like citation count and impact factor has led to inflated author lists, excessive self-citation, and weakened correlations between authorship and scholarly contribution. Their study illustrated how Goodhart’s Law manifests in academic publishing: when metrics become targets, they lose their diagnostic value.

These macro-level distortions intersect with hyperauthorship, explored by Meho \cite{meho2024using}, who documented a dramatic rise in average author counts over the past two decades. His findings show that as large-scale interdisciplinary collaborations become more common, the complexity of assigning meaningful credit increases, creating new challenges for transparency and accountability in multi-author work.

Regarding authorship ethics, Maddi and Teixeira da Silva \cite{maddi2024beyond} performed a detailed audit of 81,823 PLOS ONE articles published between 2018 and 2023. They applied formal ICMJE guidelines to detect questionable authorship practices. They found that 9.14\% of papers included at least one author whose contributions were insufficient to warrant authorship. Their analysis also identified geographic disparities, with higher rates of inappropriate authorship in certain regions, particularly among senior or industry-affiliated individuals.

Andersen and Wray addressed further ethical concerns \cite{andersen2023rethinking}, they analyzed a large corpus of retraction notices to assess how contribution statements are used—if at all—in attributing responsibility during scientific misconduct cases. Their work found that despite the availability of contribution metadata, journals and institutions rarely use it to assign accountability, suggesting a gap between the ideal of transparency and actual enforcement mechanisms.

Lastly, Patience et al. \cite{patience2019survey} conducted a broad-based survey of highly cited scientists across disciplines, querying what types of work merit authorship. The results revealed wide disciplinary variation and highlighted the need for clearer community standards. Respondents differed not only in what tasks they considered authorship-worthy, but also in their attitudes toward shared versus individual credit.

In addition to empirical and ethical analyses, several computational frameworks have been developed to support the automated processing of author contribution statements at scale. A notable early effort is NaïveRole by Tkaczyk et al.~\cite{tkaczyk2019naiverole}, which employs a rule-based classification strategy using keyword heuristics and syntactic patterns to infer author roles from unstructured text. While this approach laid foundational groundwork, its reliance on rigid pattern-matching limits its generalizability across diverse writing styles and scientific domains.

\vspace{3.5mm}
\noindent Compared to previous work, our study leverages a substantially larger and more diverse dataset, and applies a hybrid classification method tailored for unstructured text. We also introduce novel analyses—such as task co-occurrence patterns and contribution inequality modeling—while promoting transparency through public data release.

\section{Methodology}
\label{sec:methodology}
The methodology presented in this study is structured to enable a systematic analysis of author contributions at scale. Section~\ref{sec:dataset} introduces the dataset construction process, including collecting articles from two major repositories. Section~\ref{sec:preprocessing} outlines the preprocessing steps undertaken to standardize the data. This preprocessing includes author name inconsistencies, detailed in Section~\ref{sec:author_names}, and the normalization of contribution statements into a fixed taxonomy of scholarly tasks, as described in Section~\ref{sec:contrib_classification}. 

\subsection{Dataset}
\label{sec:dataset}
In this study, we focused on collecting and processing scientific articles from two major repositories--PLOS One  and Nature--to investigate patterns in author contributions. These sources provided access to a large and diverse corpus of research articles, each containing metadata fields such as publication year, list of authors, article title, and most importantly, the "Author Contributions" section.
Given the heterogeneous formats of the data sources—from online web interfaces to bulk files from journal repositories—we retrieved and parsed author contribution information. This process enabled consistent metadata representation across both sources.

We implemented a structured extraction pipeline to manage the substantial volume of collected data, storing the parsed content in a unified format. This approach enables efficient querying, transformation, and subsequent analytical processing. The final dataset included 403,000 articles containing structured metadata and a dedicated author contributions section. Across this corpus, we identified 1,582,231 author mentions, representing repeated contributions of authors across multiple articles. We mapped these contributions to 5,636,781 task assignments, as categorized by the 14 standardized CRediT roles~\cite{allen2014credit}. To the best of our knowledge, this large-scale aggregation represents one of the most comprehensive and systematically structured collections assembled for analyzing patterns in scholarly contribution at scale.

\subsection{Preprocessing}
\label{sec:preprocessing}
\subsubsection{Standardizing Author Names}
\label{sec:author_names}
A  key step in the preprocessing pipeline involved resolving inconsistencies in how authors were referred to within contribution sections. While full names are typically listed in the author metadata at the beginning of each article, contribution statements often reference contributors using initials, which vary widely in style and structure. For example, an author named ``John Doe Smith'' might appear as \textit{J.S.}, \textit{J.D.S.}, \textit{John S.}, or even \textit{Smith J.D.} within the same dataset.\footnote{
Possible variations include: 
\textit{JDS}, \textit{J.S.}, \textit{J.D.S.}, \textit{J. D. Smith}, \textit{Smith J.D.}, \textit{Smith, J.D.}, \textit{John S.}, \textit{John D. S.}, \textit{JD Smith}, \textit{J D S}, \textit{J-D-S}, \textit{jds}, \textit{j. d. smith}, \textit{smith jd}, \textit{smith john d}, \textit{jd.smith}, \textit{smith.j.d}, \textit{J D-S}, and other combinations with/without punctuation, capitalization, or ordering. These inconsistencies often include middle name abbreviations, reversed name order, hyphenated formats, and casing variations.} These variations made it challenging to accurately map initials back to full author names, which was essential for linking contributions to the correct individuals.

To address this, we systematically generated and tested all plausible combinations of initials for each author and matched them against the names mentioned in the contribution paragraph. This matching process included handling cases such as:

\begin{itemize}\setlength\itemsep{0.1em}
    \item Inverted name orders.
    \item Inclusion or exclusion of middle names.
    \item Abbreviated first names.
    \item Authors with multiple last names.
\end{itemize}

To address the variation in acronym formats, we implemented a rule-based resolution mechanism that systematically generates all plausible acronym candidates for each author. The matching process accounted for variations in name order, optional middle names, abbreviation styles, capitalization, hyphenation, and multi-part surnames. We matched each candidate acronym against the contribution statements using a dedicated function that iteratively tested combinations and applied fallback strategies based on character patterns and casing.

By successfully resolving these ambiguities, we were able to reliably associate each contribution with its corresponding author. This step was critical to enabling all subsequent analyses regarding authorship patterns, task frequencies, and role distributions. To evaluate the accuracy of this disambiguation process, we applied our algorithm to a random sample of 100 scientific articles, comprising 665 author entries. The algorithm correctly matched abbreviated initials in the contribution statements to the full names listed in the metadata with a global accuracy of \(94.73\%\), and an average per-article accuracy of \(93.87\%\). These results demonstrate that the disambiguation module performed consistently and effectively across the dataset, providing a reliable foundation for all downstream analyses (see Section~\ref{sec:results} for full evaluation details).
.

\subsubsection{Standardizing Contribution Representation}
\label{sec:contrib_classification}

To systematically analyze and compare authors' roles across a large corpus of scientific articles, we developed a structured classification pipeline for mapping free-text author contribution statements to a fixed set of standardized task categories. The goal of this pipeline was to enable large-scale, automated analysis of scholarly contribution patterns while maintaining clarity, consistency, and interpretability.

To ensure consistency and enable structured analysis, we adopted the CRediT taxonomy~\cite{allen2014credit} as the foundation for categorizing author contributions. This standardized framework defines 14 distinct roles that span research's conceptual, technical, and supervisory aspects.  These include: \textit{writing -- review editing}, \textit{methodology}, \textit{investigation}, \textit{conceptualization}, \textit{formal analysis}, \textit{data curation}, \textit{writing -- original draft}, \textit{supervision}, \textit{validation}, \textit{project administration}, \textit{resources}, \textit{funding acquisition}, \textit{visualization}, and \textit{software}. Relying on this taxonomy allowed us to map free-text contribution statements to a standard set of categories, enabling scalable and comparable analysis across a large corpus of scientific articles.

We compiled a refined list of relevant keywords and representative phrases for each category based on the PLOS One authorship guidelines\footnote{\url{https://journals.plos.org/plosone/s/authorship}} and an extensive manual review of real-world author contribution statements. The resulting lexical resources were the basis for an initial rule-based classification layer, which enabled us to match many contribution statements directly to task categories based on explicit textual cues.

However, keyword matching alone was insufficient for capturing semantically similar but lexically diverse expressions. Many valid statements lacked exact matches to the curated keywords, yet clearly aligned with one or more task categories. To address this limitation, we introduced a semantic similarity layer based on sentence embeddings and few-shot learning. For each task category, we selected 5-15 prototypical example statements reflecting typical ways that contribution to that task is phrased. These examples and category keywords were encoded using the \texttt{all-mpnet-base-v2} SentenceTransformer model \cite{reimers2020sentencebert} to produce meaningful semantic representations.

We compared statements that could not be confidently classified via keyword rules to these semantic representations using cosine similarity. Then, we assigned the statement to the category with the highest semantic alignment. Combining lexical matching, curated examples, and embedding-based semantic comparison enabled 
robust, scalable, and consistent classification across a diverse contribution landscape. We refer to this integrated rule-based and embedding-enhanced classification as \textit{H-Contrib} (Hybrid Contribution Classifier).

The following Algorithm~\ref{alg:hcontrib} outlines the hybrid classification procedure used in our pipeline

\begin{algorithm}
\caption{Hybrid Contribution Classifier (H-Contrib)}
\label{alg:hcontrib}
\begin{algorithmic}[1] 
\State \textbf{Input:} Contribution statements $S = \{s_1, ..., s_n\}$
\State \textbf{Output:} Task categories $c_i$ for each $s_i$
\State Define categories $C = \{c_1, ..., c_{14}\}$
\ForAll{$c_j \in C$}
    \State Compile keywords $K_j$
    \State Collect example phrases $E_j$
\EndFor
\ForAll{$s_i \in S$}
    \If{$s_i$ matches any $k \in K_j$}
        \State Assign $s_i \rightarrow c_j$
    \Else
        \State Encode $s_i$ as vector $v_i$
        \ForAll{$c_j \in C$}
            \State Embed category from $E_j$ to get $u_j$
        \EndFor
        \State Compute cosine similarity between $v_i$ and each $u_j$
        \State Assign $s_i \rightarrow c_k$ with the highest similarity
    \EndIf
\EndFor
\end{algorithmic}
\end{algorithm}
The H-Contrib algorithm begins by defining the complete set of task categories (line 3), each associated with a curated list of keywords and representative example phrases (lines 4–6). These serve as lexical and semantic anchors for classification.
Next, the algorithm iterates through each contribution statement (line 8). If a statement matches any predefined keyword, it is assigned directly to the corresponding task category (lines 9–10), ensuring high precision through rule-based heuristics.

The algorithm leverages semantic similarity if no lexical match is found (lines 11 onward). Each unmatched statement is embedded into a vector representation (line 12) using the SentenceTransformer model. Each category is similarly represented by the mean embedding of its examples (lines 13–14). The statement vector is then compared against each category vector using cosine similarity (line 15), and the category with the highest similarity is selected as the final assignment (line 16).

To evaluate the performance of the H-Contrib algorithm, we conducted an empirical analysis on a manually annotated sample of \( K \) scientific articles. Each article included labeled task contributions, along with associated author information.

We assessed the classification and author-matching components using two complementary metrics: \textit{global accuracy}, which captures overall correctness across the entire dataset, and \textit{average per-article accuracy}, which reflects model performance at the document level.

For task classification, the global accuracy is defined as the ratio of correctly predicted task assignments to the total number of ground-truth labels across all articles:

\begin{equation*}
\text{Accuracy}_{\text{global}} = 
\frac{\sum_{k=1}^{K} \sum_{i=1}^{n_k} \text{correct}_{k,i}}{\sum_{k=1}^{K} n_k}.
\end{equation*}

Here, \( n_k \) denotes the number of labeled tasks in article \( k \), and \( \text{correct}_{k,i} \) equals 1 if the $i^{th}$ task in article \( k \) was correctly classified, and 0 otherwise.

The average per-article accuracy computes accuracy within each article and then averages the result over the corpus, giving equal weight to all documents:

\begin{equation*}
\text{Accuracy}_{\text{avg}} = 
\frac{1}{K} \sum_{k=1}^{K} \left( \frac{\sum_{i=1}^{n_k} \text{correct}_{k,i}}{n_k} \right).
\end{equation*}

The same evaluation framework was applied to the author name disambiguation module, which links abbreviated initials in contribution statements to the corresponding full names in the article metadata. The global accuracy for author matching is given by:

\begin{equation*}
\text{Accuracy}_{\text{global}}^{\text{author match}} = 
\frac{\sum_{k=1}^{K} \sum_{i=1}^{m_k} \text{correct}_{k,i}}{\sum_{k=1}^{K} m_k},
\label{eq:author_match_global}
\end{equation*}

where \( m_k \) is the number of authors in article \( k \), and \\ 
$correct_{k,i} = 1$ if the initials were correctly matched to the full name.

The average per-article accuracy for author matching is similarly defined as:

\begin{equation*}
\text{Accuracy}_{\text{avg}}^{\text{author match}} = 
\frac{1}{K} \sum_{k=1}^{K} \left( \frac{\sum_{i=1}^{m_k} \text{correct}_{k,i}}{m_k} \right).
\label{eq:author_match_avg}
\end{equation*}

Together, these metrics offer both a corpus-wide and article-specific perspective on the accuracy of task classification and author resolution, enabling a robust system performance evaluation.

We employed the Apriori algorithm~\cite{agrawal1994fast}—a foundational method in frequent itemset mining—to detect standard sets of contribution categories reported together within the
same article to uncover recurring patterns in how contribution roles co-occur. We represented each as a transaction containing the unique contribution roles assigned to its authors. We applied a minimum support threshold of 0.2, meaning we retained only those role combinations that appeared in at least 20\% of the articles. In this context, support measures how frequently a given set of roles appears across the entire dataset, indicating how representative or common that pattern is. This threshold allowed us to focus on dominant and meaningful task bundles while filtering out infrequent or incidental co-occurrences.
To structure our empirical analysis, we formulated a series of research objectives aimed at uncovering key patterns in author contribution dynamics and evaluating the effectiveness of our proposed classification approach:
\begin{enumerate}[leftmargin=*, label=\textbf{\arabic*.}, align=left, labelsep=0.5em]
    \item \textbf{Authors’ Contribution Analysis.}  
    We aimed to examine the overall distribution of contribution types across a large-scale dataset of scientific articles. This analysis involved quantifying how frequently each of the 14 CRediT-defined roles appeared across millions of annotated author-task assignments. In addition, we applied the Apriori algorithm to detect frequent co-occurrence patterns among contribution types. Using the Apriori algorithm enabled us to uncover task bundles that commonly appear together in author statements, offering deeper insight into how scholarly responsibilities are structured and shared within collaborative research teams.

    \item \textbf{Authors’ Position in the Author List.}  
   We analyzed how the relative contribution of authors varies with their position in the byline across different team sizes. Specifically, for each author position (e.g., position 4), we computed the average share of contributions that authors in that position received across all articles in which the total number of authors was greater than or equal to their position (e.g., 4 to 20 authors). This allowed us to assess how the expected contribution for each author position evolves across varying collaboration sizes, providing a systematic view of positional patterns in multi-author teams.

    To further analyze this relationship, we examined how the average relative contribution of selected author positions changes as a function of team size. For this, we fixed several representative positions (e.g., author \#1, \#5, \#10, \#15) and tracked their average task share across articles with increasing numbers of authors. This detailed resolution enables us to explore the impact of team expansion on individual authors' visibility and involvement within collaborative publications.

    \item \textbf{Connection Between Author Position and Task Type.}  
    We examined whether specific contribution categories— such as supervision, funding acquisition, formal analysis, or software—are more likely to be associated with particular author positions. This analysis was designed to identify structural patterns in how academic roles are distributed within collaborative teams.

    \item \textbf{Predicting Contribution Roles.}  
    We developed a supervised multi-label classification model to predict the contribution types associated with each author. The prediction task was framed as a multi-label classification problem, where the \textit{dependent variables} were the set of CRediT contribution categories assigned to each author. The \textit{independent variables} consisted of the following metadata-based features: the author's position within the byline, the number of tasks assigned to the author, the total number of co-authors on the article, the total number of tasks performed by all authors in the article, and the author's relative share of contribution tasks within the article. 
    
    To evaluate model performance, we compared the classification accuracy of three approaches: an XGBoost model~\cite{chen2016xgboost}, a logistic regression model~\cite{cox1958regression}, and a baseline dummy classifier that always predicts the most frequent label. This comparison allowed us to assess the models’ effectiveness in capturing contribution-label patterns across heterogeneous scientific contexts.

\end{enumerate}
\section{Results}
\label{sec:results}
We executed our methodology on the constructed dataset and report the key findings below.

\noindent To evaluate the effectiveness of our classification algorithm, we applied it to a random sample of \( K = 100 \) scientific articles. This subset included a total of \( \sum_{k=1}^{K} n_k = 505 \) manually verified task assignments, where \( n_k \) denotes the number of task assignments in article \( k \). We compared each prediction produced by the model against the ground-truth labels extracted from the original author contribution statements. Based on this evaluation, the global accuracy (\( \text{Acc}_{\text{global}} \)) was \( 93.81\% \), and the average per-article accuracy (\( \text{Acc}_{\text{avg}} \)) was \( 92.24\% \).

\noindent To evaluate the effectiveness of our author name disambiguation module, we conducted a parallel assessment on the same random sample of \( K = 100 \) scientific articles. This subset included a total of \( \sum_{k=1}^{K} m_k = 665 \) author entries, where \( m_k \) denotes the number of authors in article \( k \). For each author, we evaluated whether the algorithm correctly matched the abbreviated initials appearing in the contribution statements to the corresponding full name listed in the metadata. Based on this evaluation, the global author matching accuracy (\( \text{Acc}_{\text{global}}^{\text{author match}} \)) was \( 94.73\% \), and the average per-article author matching accuracy (\( \text{Acc}_{\text{avg}}^{\text{author match}} \)) was \( 93.87\% \).
\vspace{1em}

\begin{figure}[htbp]
    \centering   \includegraphics[width=1\linewidth, height=1\textheight, keepaspectratio]{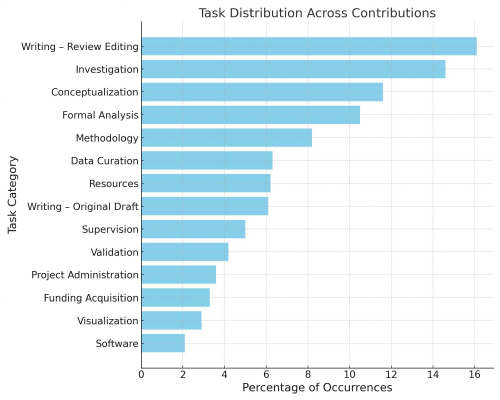}
    \caption{Distribution of author contributions types.}
    \label{fig:fig2}
\end{figure}

\noindent Following this evaluation, we analyzed the whole dataset to understand the broader distribution of contributions. As shown in Figure~\ref{fig:fig2}, \textit{writing – review editing} emerged as the most frequently reported task, while \textit{software development} was the least common.

\noindent To uncover how contributions are commonly grouped together, we applied the Apriori algorithm~\cite{agrawal1994fast} to identify frequent co-occurrence patterns. Table~\ref{tab:apriori_top_pairs} lists six dominant pairings (Support $\geq 20\%$), with \textit{writing – review editing} frequently co-occurring with tasks such as \textit{investigation} and \textit{conceptualization}.

\begin{table}[ht]\scriptsize
\renewcommand{\arraystretch}{1.4} 
\centering
\caption{Top Frequent Task Pairings (Support $\geq 20\%$).}
\label{tab:apriori_top_pairs}
\begin{tabular}{@{}lc@{}}
\toprule
\textbf{Task Pair} & \textbf{Support (\%)} \\
\midrule
Investigation $\rightarrow$ Writing–Review Editing & 30.40 \\
Writing–Review Editing $\rightarrow$ Conceptualization & 29.43 \\
Formal Analysis $\rightarrow$ Writing–Review Editing & 27.74 \\
Formal Analysis $\rightarrow$ Investigation & 22.59 \\
Investigation $\rightarrow$ Conceptualization & 22.33 \\
Formal Analysis $\rightarrow$ Conceptualization & 21.70 \\
\bottomrule
\end{tabular}
\end{table}

As shown in Table~\ref{tab:apriori_top_pairs}, several task combinations appear together in a large portion of the data. The most frequent pairings involve tasks such as \textit{writing – review editing}, \textit{investigation}, \textit{formal analysis}, and \textit{conceptualization}, which often co-occur within the same contribution statements. These combinations consistently exhibit support levels exceeding 20\%, indicating that these roles are commonly performed in conjunction with one another across a wide range of author profiles.

\begin{figure}[htbp]
    \centering
    \includegraphics[width=1\linewidth, height=1.05\textheight, keepaspectratio]{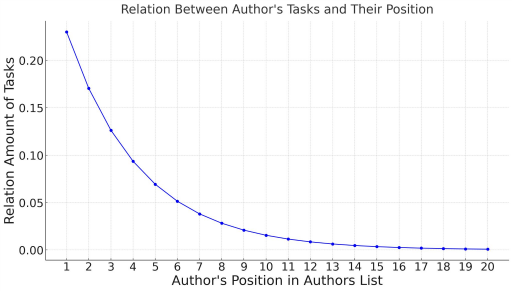}
    \caption{The average number of related tasks per author position is in the byline.}
    \label{fig:authors_fig}
\end{figure}

We next explored how contribution levels vary as a function of an author’s position in the byline. As illustrated in Figure~\ref{fig:authors_fig}, authors listed earlier—particularly those in the first or second position—tend to report a greater number of distinct contributions. In contrast, later authors are associated with fewer tasks on average. The trend demonstrates  an apparent and steep decline in contribution density, with the first author accounting for more than 20\% of the total task assignments. From the third author onward, the average share of contributions drops sharply. It continues to taper off gradually, reaching very low levels beyond the $15^{th}$ author.

To extend our investigation, we examined how authors’ contributions vary not only by their position in the author list, but also in relation to the total number of authors included in a publication. This analysis aimed to determine whether a systematic pattern exists in the distribution of contribution percentages relative to an author’s position and the overall group size.
\begin{figure}[htbp]
    \centering
    \includegraphics[width=1\linewidth, height=1\textheight, keepaspectratio]{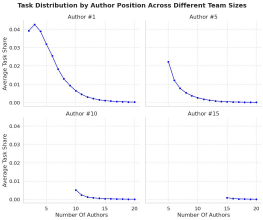}
    \caption{The connection between the authors' position in the author list and the 
relation average tasks the author performed across varying author counts.}
    \label{fig:Task Distribution Author Position}
\end{figure}

Figure~\ref{fig:Task Distribution Author Position} reveals several notable patterns. The first author consistently contributes the highest task share, peaking at approximately 4.3\% in three-author teams and gradually declining to below 0.3\% in twenty-author teams. In contrast, mid- and late-positioned authors, such as positions 10 and 15, contribute significantly less, often under 0.03\% and 0.01\%, respectively. These figures highlight the sharp concentration of contributions among early-listed authors in larger teams.

\begin{figure}[htbp]
    \centering
    \includegraphics[width=1\linewidth, height=1\textheight, keepaspectratio]{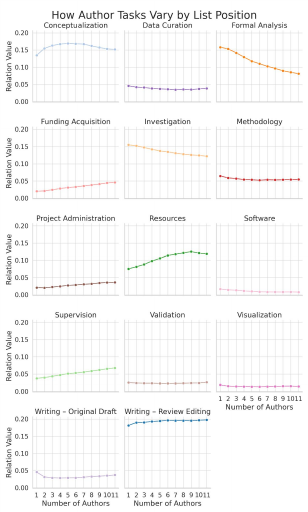}
    \caption{Relationship between task type and author position in the byline. Each subplot shows how often a specific contribution type is linked to different author positions. Tasks are sorted by their maximum relative contribution to highlight structural differences across roles.}
    \label{fig:authors place to tasks}
\end{figure}

\begin{figure}[htbp]
    \centering
    \includegraphics[width=1\linewidth, height=1\textheight, keepaspectratio]{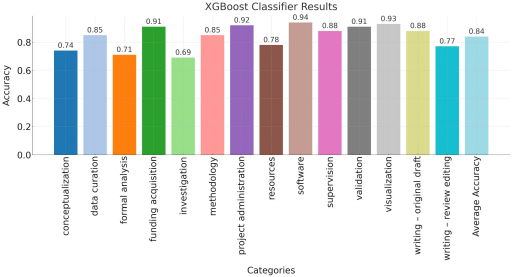}
    \caption{XGBoost classifier model results for each contribution type task. 
    The model with the best performance.}
    \label{fig:ModelResults}
\end{figure}

We examined whether certain types of contributions are more commonly associated with specific author positions.
The results, shown in Figure~\ref{fig:authors place to tasks}, reveal that \textit{writing – review editing} is the most frequently reported task across nearly all author positions, maintaining high relation values even as the number of authors increases.
Tasks such as \textit{supervision} and \textit{funding acquisition} demonstrate a gradual increase with author position, peaking toward the later positions in the byline.
Conversely, technical and hands-on tasks such as \textit{software}, \textit{investigation}, and \textit{formal analysis} exhibit a noticeable decrease as the author's position increases.
Tasks like \textit{project administration} and \textit{resources} show a moderate rise, while \textit{validation} and \textit{data curation} remain relatively stable across author positions and team sizes.

Lastly, we evaluated the predictive performance of our classification models across the 14 standardized contribution categories. As shown in Figure~\ref{fig:ModelResults}, the XGBoost classifier achieved the highest overall accuracy, outperforming both logistic regression and a baseline dummy classifier. This result indicates that metadata-based features such as author position and team composition carry strong predictive signals for contribution role classification.

Lastly, we evaluated the predictive performance of three classification models—XGBoost, logistic regression, and a baseline DummyClassifier—across the 14 standardized contribution categories. As shown in Figure~\ref{fig:ModelResults}, the XGBoost model achieved the highest average accuracy at 0.84, followed by logistic regression at 0.81, and the Dummy baseline at 0.75.

Performance varied across contribution types. The model showed particularly strong results for specialized roles such as \textit{software}, \textit{visualization}, and \textit{validation}, which tend to exhibit distinct statistical patterns. In contrast, more general-purpose categories like \textit{writing – review editing} and \textit{conceptualization} yielded lower predictive accuracy, likely due to their broader and more uniform distribution across authorship positions.
.

A detailed view of the XGBoost model’s predictive behavior is presented in the confusion matrices presented in the Appendix (Figure~\ref{fig::conf_matrix_all}. These visualizations illustrate the per-category classification performance, highlighting the strengths and the common sources of misclassification across task types.

\section{Discussion}
\label{sec:discussion}

Our analysis provides several key insights into scholarly contributions' structure, distribution, and predictability within large-scale collaborations.

First, we can observe the performance of the H-Contrib classification algorithm from our results. Applied to a manually annotated sample of \( K = 100 \) scientific articles encompassing a total of \( \sum_{k=1}^{K} n_k = 505 \) assignment of contribution tasks - where \( n_k \) denotes the number of contribution tasks assigned in an article \( k \) - the algorithm achieved a global precision of \( 93.81\% \) and an average accuracy per article of \( 92.24\% \). These figures indicate that our hybrid rule-based and semantic similarity techniques can reliably map free-text contribution statements into structured categories.
In parallel, we evaluated the accuracy of our author name disambiguation module using the same sample of \( K = 100 \) articles, which included a total of \( \sum_{k=1}^{K} m_k = 665 \) authors, where \( m_k \) represents the number of authors in article \( k \). This component achieved a global accuracy of \( 94.73\% \) and an average per-article accuracy of \( 93.87\% \) in correctly linking abbreviated initials to full author names listed in the metadata. These results demonstrate the robustness of our matching procedure and its effectiveness in resolving ambiguous or abbreviated author references, a vital step in ensuring the validity of our analyses.

In parallel, we evaluated the accuracy of our author name disambiguation module using the same sample of \( K = 100 \) articles, which included a total of \( \sum_{k=1}^{K} m_k = 665 \) authors. This component achieved a global accuracy of \( 94.73\% \) and an average per-article accuracy of \( 93.87\% \) in correctly linking abbreviated initials to full author names listed in the metadata. These results demonstrate the robustness of our matching procedure and its effectiveness in resolving ambiguous or abbreviated author references, a vital step in ensuring the validity of our analyses.

Prior studies have made substantial contributions to understanding how scientific labor is distributed among co-authors. For example, Corrêa Jr.~\cite{correa2017patterns}, Macaluso et al.~\cite{macaluso2016gender}, and Sauermann and Haeussler~\cite{sauermann2017authorship} examined authorship practices using manual annotations and survey-based insights, shedding light on role patterns, gender disparities, and the inconsistencies between author order and actual work performed. Building on this empirical foundation, Larivière et al.~\cite{lariviere2021division} and Maddi and Teixeira da Silva~\cite{maddi2024beyond} utilized structured metadata (e.g., CRediT) to analyze task distributions and detect inappropriate authorship practices at scale.
Our work extends these contributions in several key ways. First, it operates on over 337,000 full-text publications, encompassing more than 5.6 million author task associations, an order of magnitude larger than previous datasets. Second, rather than relying on structured metadata, we analyze free-text statements using a hybrid classification framework combining rule-based logic and semantic similarity. Our method enables broader applicability across diverse writing styles and publication venues. Third, our application of association rule mining (Apriori) allows us to model task co-occurrence patterns within articles, going beyond frequency-based metrics to reveal structural relationships among roles. Fourth, we quantify contribution inequality as a function of team size, an analysis that has not, to our knowledge, been formalized in prior studies. Finally, we release our full dataset for public use, supporting open science and reproducibility.

Second, we can observe a key insight that involves the distribution of contribution types across different author positions. 
 Analysis of the full dataset revealed that \textit{writing – review editing} was the most frequently reported task across nearly all author positions (Figure~\ref{fig:fig2}). This result suggests that manuscript reviewing and editing are widely shared responsibilities in scientific collaborations. However, the frequent co-occurrence of \textit{writing – review editing} with a broad range of other tasks (Table~\ref{tab:apriori_top_pairs}) may reflect varying degrees of involvement in the writing process, ranging from substantive drafting and revision to later-stage editorial input.

Third, our analysis utilizing the Apriori algorithm revealed interesting patterns of task co-occurrence, providing a deeper understanding of how research activities are combined within collaborative projects. As shown in Table~\ref{tab:apriori_top_pairs}, the most frequent pairing was between \textit{investigation} and \textit{writing – review editing}, with a support level of 30.4\%. This result suggests that authors involved in the core investigation phase of research often also take part in reviewing and refining the manuscript. Other strong associations we uncovered included \textit{writing – review editing} co-occurring with \textit{conceptualization} (29.43\%) and \textit{formal analysis} with both \textit{writing – review editing} (27.74\%) and \textit{investigation} (22.59\%). These findings point to a structural pattern where intellectual tasks—particularly the generation of ideas, data analysis, and primary investigation—are tightly coupled with responsibilities related to manuscript development and refinement.

Notably, the presence of multiple high-support pairings involving \textit{writing – review editing} highlights the role of writing as a collaborative process intertwined with various stages of scientific work, rather than a distinct, isolated activity. Instead of authors specializing narrowly in only technical or only editorial tasks, our findings suggest that substantial overlaps exist, reflecting the integrated nature of modern collaborative research efforts. The frequent coupling of technical and editorial contributions emphasizes that the major contributors are also frequently involved in presenting and communicating scientific findings.

Fourth, we observed consistent positional trends linking author order to task types. As shown in Figure~\ref{fig:authors_fig}, authors listed earlier in the byline tend to report a greater number of distinct contributions compared to those listed later. In particular, technical tasks like \textit{formal analysis}, \textit{investigation}, and \textit{software}  were primarily associated with early-positioned authors, while leadership-oriented tasks such as \textit{supervision} and \textit{funding acquisition} became more common toward the end of the byline (Figure~\ref{fig:authors place to tasks}). For example, the rate of supervision among authors ranked eighth and onward was more than twice that of first-position authors. Such positional regularities could potentially be leveraged for anomaly detection, identifying cases where an author's task profile deviates from expected patterns

Fifth, we found that team size significantly impacts contribution inequality. As demonstrated in Figure~\ref{fig:min max ratio tasks}, small teams displayed relatively balanced task participation, with an average ratio of 2.3 between the most and least active contributors in two-author teams. In contrast, in collaborations involving 20 authors, this disparity widened to a ratio of 4.7, reflecting the increased concentration of work among core team members as collaboration size grows. 

To quantify this relationship, we derived a linear model: 
\(\mathrm{Ratio}_{\mathrm{max/min}} = 3.000 + 0.082 \cdot \mathrm{TeamSize}\),
indicating that with each additional team member, the gap between the most and least active contributors grows on average by approximately 0.082.
This linear trend suggests that contribution imbalance is not merely a sporadic phenomenon in large teams, but a systematic structural feature that scales with team size. As collaborations expand, a core-periphery dynamic becomes more pronounced, with a smaller subset of authors bearing a disproportionately larger share of the overall workload. Understanding this dynamic is crucial for promoting fair recognition practices in multi-author publications, especially in disciplines where hyperauthorship is becoming increasingly common.

Sixth, examining task-specific trends along the byline revealed that contributions such as \textit{conceptualization} and \textit{resources} were more common among early and middle authors. In contrast, supervision and \textit{funding acquisition} were increasingly reported by later authors. Early-positioned contributors almost exclusively performed technical tasks like \textit{software} and \textit{formal analysis}. These findings reinforce the functional meaning embedded in author order conventions.

Lastly, our predictive models demonstrated that they can effectively infer the contribution types. The XGBoost classifier achieved an average accuracy of \( 84\% \) across the 14 contribution categories, outperforming logistic regression (\( 81\% \)) and a Dummy baseline classifier (\( 75\% \)) (Figure~\ref{fig:ModelResults}). Specialized contributions such as \textit{software}, \textit{visualization}, and \textit{validation} were predicted with high accuracy. In contrast, broader, more diffuse tasks like \textit{writing} and \textit{conceptualization} proved more challenging to classify.

Despite these contributions, several limitations of our study should be acknowledged. 

First, although the dataset analyzed in this research is substantial—comprising more than 337,000 scientific articles and approximately 1.5 million author mentions—it was limited to two repositories: journals from the Nature Publishing Group and PLOS One. Although these platforms are widely respected and provide structured author contribution statements, their inclusion may introduce disciplinary or institutional biases. The findings may not generalize to fields or journals that lack formalized contribution reporting, such as computer science conferences or humanities journals. Furthermore, variation in editorial policies across even structured platforms may influence the phrasing, granularity, and completeness of contribution descriptions.

Second, despite the robustness of our classification pipeline, free-text contribution statements remain inherently ambiguous and inconsistent. Authors may describe the same activity using semantically distinct terms or use identical terms to refer to different types of work, depending on context. While our hybrid method—combining rule-based heuristics with semantic similarity—mitigates this challenge, classification errors may persist, particularly for vague or underspecified statements. This issue can be further compounded by interdisciplinary terminology drift, where technical terms carry different meanings across fields ~\cite{zhang2023semanticdrift}.

Third, our predictive modeling of author roles, based solely on metadata features such as author position and team size, achieves good performance (accuracy 0.84). However, such models risk reinforcing existing positional biases. They cannot account for less visible contributions (e.g., mentorship, ideation) not captured by structured metadata. In other words, while position is strongly correlated with task type, it is not a perfect proxy for actual contribution behavior.

\section{Conclusions and Future Work}
\label{sec:conclusion}
In this study, we conducted a large-scale analysis of author contributions across 403,000 scientific articles and 5,636,781 task assignments. We used a hybrid classification pipeline to transform unstructured free-text statements into standardized CRediT contribution categories. Our work sheds light on how credit is distributed in scholarly collaborations, particularly regarding author position and team size.

We found that writing-related tasks are widely shared among contributors.  At the same time, technical roles such as software development and formal analysis are concentrated among early-positioned authors. Senior authors, especially those listed last, are frequently involved in project leadership and funding-related activities. These patterns align with established academic norms, but our analysis provides empirical validation at scale. We also demonstrated that contribution roles can be predicted with high accuracy based on simple metadata features, achieving 0.84 average accuracy across 14 task types using an XGBoost classifier. This result suggests the feasibility of automating contribution inference to support authorship transparency and fair credit attribution.

Our findings also reveal significant disparities in the number of tasks authors perform based on their position in the byline. Early-positioned authors tend to carry out a larger number of distinct contributions than middle- and late-positioned authors. This imbalance becomes more pronounced as team size increases, highlighting the emergence of a core-periphery structure within extensive collaborations. Such patterns emphasize the importance of considering contribution type and volume when assessing individual contributions in multi-author research.

Our study opens several avenues for future research. First, it would be valuable to examine how the distribution of author contributions evolves, revealing potential shifts in collaboration practices across different periods. Second, expanding the analysis across various disciplines could uncover domain-specific authorship behavior and contribution assignment patterns. Third, identifying anomalous cases—such as authors occupying lower-ranked positions in the byline while performing a disproportionately high number of tasks—could provide new insights into credit allocation dynamics and potential deviations from conventional authorship norms. Investigating these directions would contribute to a deeper understanding of the complexities and evolution of scientific collaboration.

By formalizing and quantifying the distribution of scholarly labor, our work contributes to building transparent, equitable, and data-driven frameworks for academic recognition.

\section*{Data and Code Availability}
\addcontentsline{toc}{section}{Data and Code Availability}

\begin{sloppypar}
Structured data generated in this study—including mappings between articles and author order, as well as between articles, authors, and their assigned contribution roles—is available at the following \href{https://drive.google.com/drive/folders/1qx99gcCkwk8BiTf4huqAZHWuwfg__nY9?usp=sharing}{link}.
\end{sloppypar}

\begin{sloppypar}
The code used for data preprocessing, classification, and analysis is available here:
\href{https://github.com/itaiassraf/Big-Data-Mining-Project}.
\end{sloppypar}

\section*{Authors' Contributions}
\addcontentsline{toc}{section}{Authors' Contributions}

M.F. conceived the idea and supervised the project.  
I.A. implemented the software, constructed the datasets used in the project, and created the visualizations.  
Both authors designed and conducted the experiments, and contributed to writing and editing the manuscript.  

All code, data, and arguments were iterated over multiple times — including late-night commits, caffeinated debates, and at least one existential crisis over a mislabeled task category.

\section*{Acknowledgment}
\addcontentsline{toc}{section}{Acknowledgment}
The authors acknowledge the assistance of AI tools, such as ChatGPT and Grammarly, in improving and editing the manuscript. 


\bibliographystyle{unsrt}
\bibliography{Behind_the_Byline_A_Large_Scale_Study_of_Scientific_Author_Contributions}

\clearpage

\appendix
\renewcommand{\thefigure}{S\arabic{figure}}
\renewcommand{\thetable}{S\arabic{table}}
\setcounter{figure}{0}
\setcounter{table}{0}
\section*{Appendix}
\addcontentsline{toc}{section}{Appendix}
\begin{table}[h!]
\centering
\small
\caption{Clean and compact keyword mapping for each contribution category. Constructed based on the CRediT taxonomy and the PLOS One guidelines.}
\caption*{\href{https://journals.plos.org/plosone/s/authorship}{link}}
\renewcommand{\arraystretch}{1.6}
\rowcolors{2}{gray!10}{white}
\begin{tabular}{>{\raggedright\arraybackslash}p{0.23\linewidth} >{\raggedright\arraybackslash}p{0.23\linewidth}
                >{\raggedright\arraybackslash}p{0.23\linewidth} >{\raggedright\arraybackslash}p{0.23\linewidth}}
\rowcolor{gray!25}
\textbf{Category} & \textbf{Keywords} & \textbf{Category} & \textbf{Keywords} \\

Writing – Review Editing & manuscript, final version, paper, publish, literature, approval, revision, review, edit, figures, proofreading, article, discussion, writing &
Methodology & model, methodology, algorithm \\

Investigation & experiment, patient, simulation, field work, fieldwork, investigation & Conceptuali-
zation &  concept, idea, initiation, conceived, study design \\

Formal Analysis & analyze, analysis, computational, interpretation, statistical, mathematical &
Data Curation & collection, generation, preparation, curation, extraction, integration, acquisition, contribution, gathering, database, cleaning, management, compilation \\

Writing – Original Draft & draft &
Supervision & supervision \\

Validation & validation, verification, replication, reproduction &
Project Administration & administration, guidance, coordination, leadership, technical, logistics, management \\

Resources & materials, reagents, tools, permission, resources &
Funding Acquisition & financing, funding, money, acquisition \\

Visualization & visualization, graph, figure, diagram, chart &
Software & software, programming, code, coding, implementation \\
\end{tabular}
\label{tab:clean-keyword-map}
\end{table}

\begin{figure*}[t]
    \centering
    \vspace{0.5em}

    \begin{minipage}[b]{0.23\textwidth}
        \includegraphics[width=\linewidth]{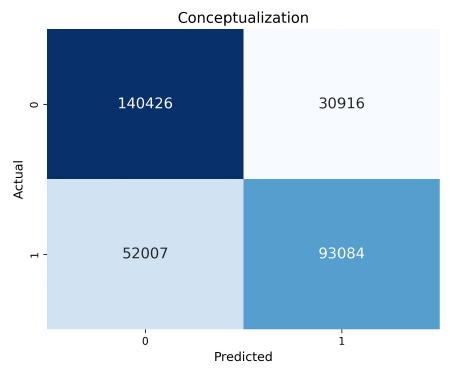}
        \caption*{Conceptualization}
    \end{minipage}
    \hfill
    \begin{minipage}[b]{0.23\textwidth}
        \includegraphics[width=\linewidth]{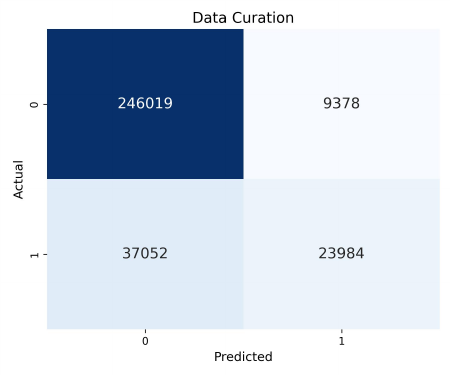}
        \caption*{Data Curation}
    \end{minipage}
    \hfill
    \begin{minipage}[b]{0.23\textwidth}
        \includegraphics[width=\linewidth]{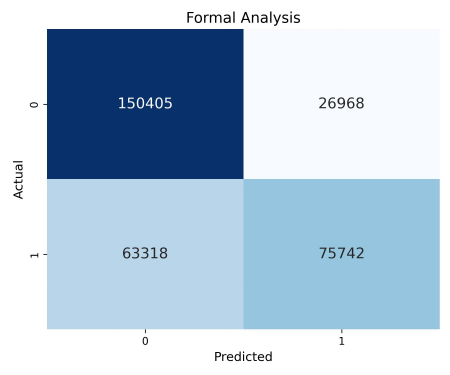}
        \caption*{Formal Analysis}
    \end{minipage}
    \hfill
    \begin{minipage}[b]{0.23\textwidth}
        \includegraphics[width=\linewidth]{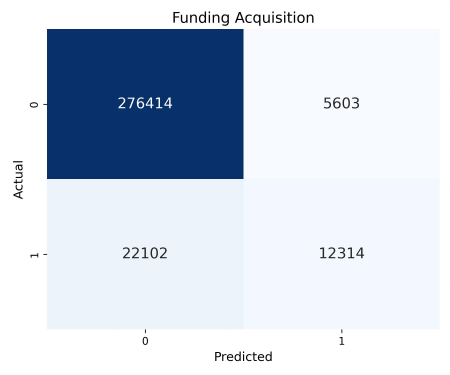}
        \caption*{Funding Acquisition}
    \end{minipage}

    \vspace{0.5em}

    \begin{minipage}[b]{0.23\textwidth}
        \includegraphics[width=\linewidth]{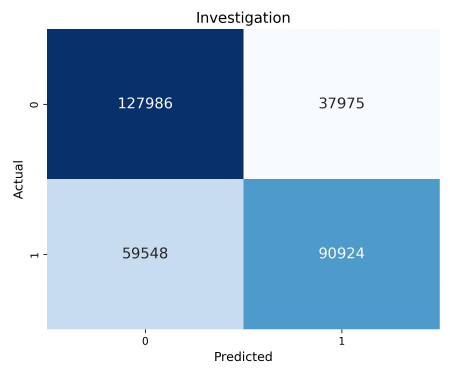}
        \caption*{Investigation}
    \end{minipage}
    \hfill
    \begin{minipage}[b]{0.23\textwidth}
        \includegraphics[width=\linewidth]{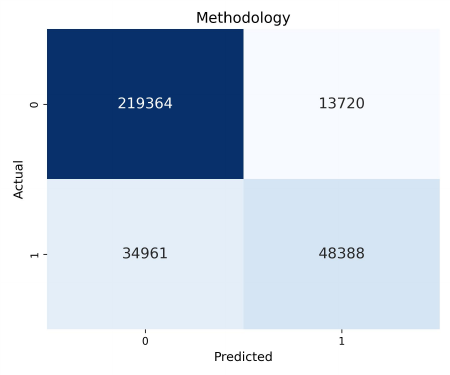}
        \caption*{Methodology}
    \end{minipage}
    \hfill
    \begin{minipage}[b]{0.23\textwidth}
        \includegraphics[width=\linewidth]{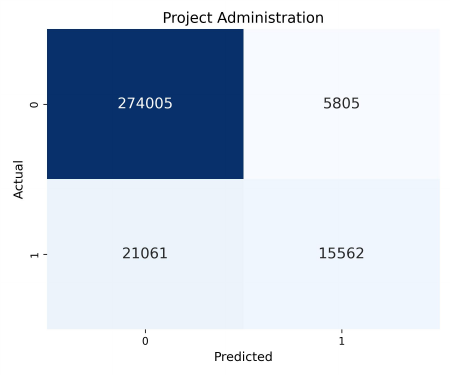}
        \caption*{Project Administration}
    \end{minipage}
    \hfill
    \begin{minipage}[b]{0.23\textwidth}
        \includegraphics[width=\linewidth]{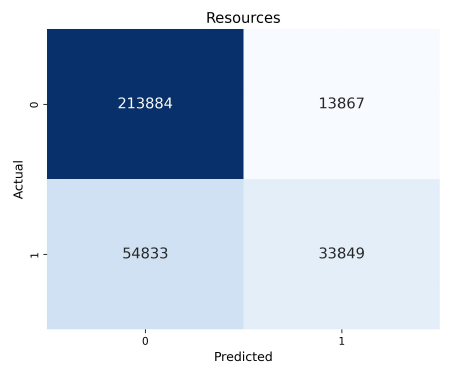}
        \caption*{Resources}
    \end{minipage}

    \vspace{0.5em}

    \begin{minipage}[b]{0.23\textwidth}
        \includegraphics[width=\linewidth]{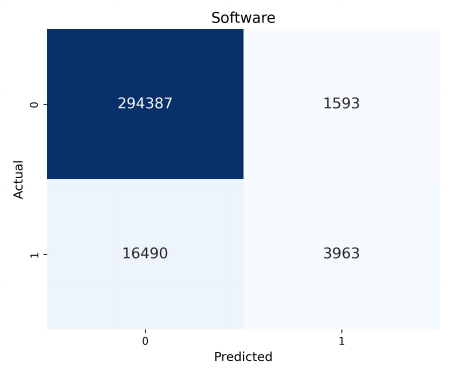}
        \caption*{Software}
    \end{minipage}
    \hfill
    \begin{minipage}[b]{0.23\textwidth}
        \includegraphics[width=\linewidth]{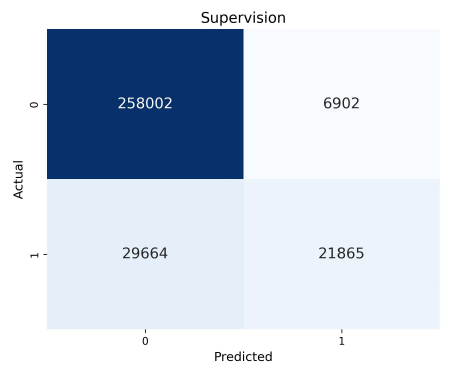}
        \caption*{Supervision}
    \end{minipage}
    \hfill
    \begin{minipage}[b]{0.23\textwidth}
        \includegraphics[width=\linewidth]{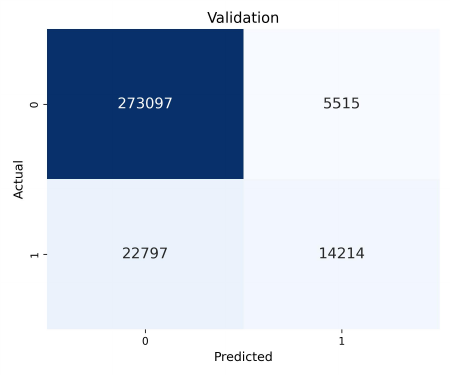}
        \caption*{Validation}
    \end{minipage}
    \hfill
    \begin{minipage}[b]{0.23\textwidth}
        \includegraphics[width=\linewidth]{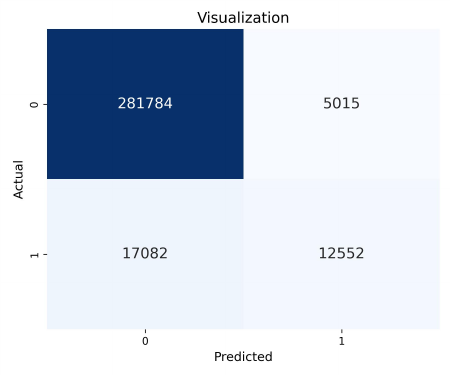}
        \caption*{Visualization}
    \end{minipage}

    \vspace{0.5em}

    \begin{minipage}[b]{0.23\textwidth}
        \includegraphics[width=\linewidth]{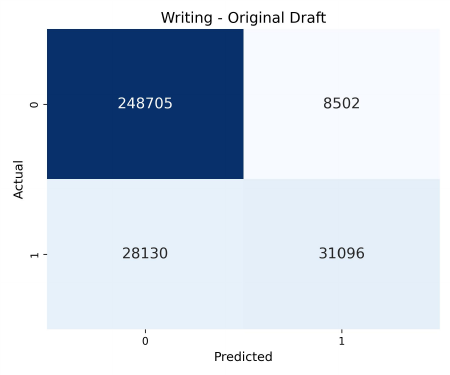}
        \caption*{Writing – Original Draft}
    \end{minipage}
    \hfill
    \begin{minipage}[b]{0.23\textwidth}
        \includegraphics[width=\linewidth]{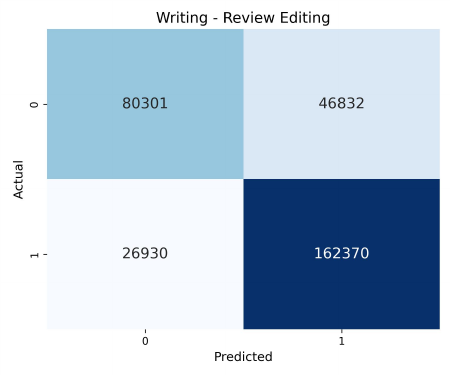}
        \caption*{Writing – Review Editing}
    \end{minipage}

\caption{Confusion Matrices for the 14 CRediT Contribution Categories}
\label{fig::conf_matrix_all}
\end{figure*}

\end{document}